*Hanrahan, B., Ning, M., and Chien Wen, Y. (2017): The Roots of Bias on Uber. In: Proceedings of 15th European Conference on Computer-Supported Cooperative Work - Exploratory Papers, Reports of the European Society for Socially Embedded Technologies (ISSN 2510-2591), DOI: 10.18420/ecscw2017-27*

# The Roots of Bias on Uber

Benjamin V. Hanrahan, Ning F. Ma, Chien Wen Yuan
College of Information Sciences and Technology at the Pennsylvania State university, USA
*bvh10@ist.psu.edu, nzm37@ist.psu.edu, tuy11@psu.edu*

**Abstract.** In the last decade, there has been a growth in, what we call, *digitally mediated workplaces*. A *digitally mediated workplace* is one where interactions between stakeholders are primarily managed by proprietary, algorithmically managed digital platform. The replacement of the relationships between the stakeholders by the platform is a key feature of these workplaces, and is a contributing factor to the decrease in contractual responsibilities each stakeholder has to one another. In this paper, we discuss some of the ways in which this structure and lack of accountability serves as a root of, or at least an enabler to, the realization of biases in the ridesharing application Uber, a digitally mediated workplace.

# Introduction

Recently, the use of *digitally mediated workplaces* has grown both in the number of participants and in the number of domains covered. Digitally mediated workplaces are primarily defined by the common stakeholder structure that they rely on, which includes: a *platform owner*, who is responsible not only for defining and implementing the platform's functionality, but also the policies around the workplace that the platform instantiates or supplements; a *worker*, who uses the platform to find, claim, and obtain remuneration from labor; and a *client*, who uses the platform to procure and pay for labor. This structure is instantiated by a number of different platforms, for a number of different purposes, e.g.: Amazon Mechanical Turk (AMT), where the worker is part of the crowd and the client requests, often small, tasks from the crowd; Fiverr, Upwork, or TaskRabbit, which are primarily aimed at freelancers to sell their services to clients; and Ola or Uber (the focus of this paper), where the worker is the driver and the client is the passenger.

The other defining feature of a digitally mediated workplace is that the – usually proprietary – platform (e.g. AMT, Fiverr, Ola, TaskRabbit, or Uber) replaces much of the relationship between the worker and the client, or the worker and employer. This has the effect to drastically alter, if not eradicate, the contractual responsibilities of each stakeholder to each other and to reduce the level of accountability all around. Aspects of this reduction of accountability have been discussed in regards to the algorithms that support the workplace in terms of algorithmic accountability (Lustig et al., 2016; Lee et al., 2015; Wagenknecht et al., 2016). However, the causes for the lack of accountability stretch beyond algorithmic mediation into the stakeholder structure and surrounding policies as well. In this paper, we explore a particular aspect of accountability in the workplace, protection from bias for both the worker and the client.

Research has already provided evidence that bias and discrimination are having a demonstrable impact on the participants of these platforms (Hannák et al., 2017; Edelman and Luca, 2014). However, this work has looked more at proving the existence of bias and less about how biased decisions are performed on or via these platforms. As we begin to investigate how we might design platforms that better support a more equitable and fair digitally mediated workplace, we first need to understand how bias is specifically occurring and what the roots of these practices might be on a specific platform. In this paper, we report on Uber, a ridesharing application where passengers obtain rides from independent drivers that use their own cars.

While Uber is not a wholly digital workplace, we argue that it is a digitally mediated one. That is, Uber provides an interesting case where there are face-to-face interactions between the driver and the passenger, these exchanges are arranged via the Uber app, and the consequences of the interactions are mediated by the app. In this way, Uber serves as an interesting and complex mixed-setting for a digitally mediated workplace, as consequences of the face-to-face interactions are both captured and propagated via the digital platform.

This is a particularly interesting setting to examine how bias functions in a digitally mediated workplace since the face-to-face interactions are arranged and the subsequent ratings of the interaction mediated solely through digital means. Meaning that, there is no human in the loop to take different factors into account or impart a level of flexibility or subjectivity to the process. As these ratings have a real impact on both the driver and passenger's ability to provide and procure services, this opens up an avenue for unfettered biased judgements that are propagated by the platform (Mcgregor et al.). To best illustrate our point, we provide this speculative comparison. In an existing, more traditional taxi service, if a passenger would like to make a biased complaint they must call a supervisor, or at the very least a representative of the taxi service. During this call, there is a likelihood that the supervisor may uncover or detect the bias due to the existing relationship between the supervisor and the driver, in addition to the supervisor's judgement as to the validity and veracity of the complaint. So, there is at least some level of human mediation when fielding complaints. Contrast this to a biased complaint on Uber, where the only signal of the complaint is a rating, which is stripped of all the nuance and reasoning behind the decision. This biased judgement is then propagated by the system, as that biased rating is used by the system and its users to determine which driver to select for a ride.

In this paper, we draw from a similar method used by Martin et al. (2014), where we examine what discussions Uber drivers are having regarding bias online. We argue, similarly to Martin et al. (2014), that Uber, like Amazon Mechanical Turk, is a digitally mediated workplace and that online forums are a place where the *shop talk* happens. When we set out to study these forums, we were interested in the social dynamics – as perceived by the drivers – that revolved around driver/passenger interactions. When we encountered posts by drivers discussing bias, it quickly became a topic of interest based on both the data and previous literature (Rosenblat et al., 2016; Nardi, 2015; Rogers, 2015; Mcgregor et al.; Raval and Dourish, 2016; Glöss et al., 2016). While we were not surprised to find that drivers discussed biases directed at them, we were surprised that they also discussed the types of biases that they had developed while driving for Uber. In this paper, we report some of our preliminary findings on how biases bear out both *by and towards* drivers on Uber and the role of the platform. In this way, we begin to look at how the same phenomena that led to protections for workers and customers in traditional workplaces are reoccurring in digitally mediated ones. The first step in dealing with bias in a computer system is to analyze its practice (Friedman, 1996), therefore our analysis of the practice of bias is the first step towards designing more equitable and fair digitally mediated workplaces. This topic is of particular importance to Computer Supported Cooperative Work (CSCW), as the 'Computer Supported' part of CSCW becomes even more consequential to the supported work, when the work is primarily instantiated and mediated by a digital platform.

# Related Work

In this section, we review research into digital mediation of work and how biases may be enacted in a digitally mediated workplace.

### Peer-to-peer platforms and technological mediation

Beyond the more traditional CSCW tools that mediate work, e.g. email (Hinds and Kiesler, 1995), instant messaging (Isaacs et al., 2002), or social network sites (DiMicco et al., 2008), peer-to-peer (P2P) platforms such as Uber, Lyft, or Ola are digitally mediated workplaces where workers manage their tasks and negotiate transactions with their clients both online and offline. While on some platforms the task may be completed offline, such as driving the passenger to a destination and potentially engaging in social interactions with each other along the way (Raval and Dourish, 2016; Glöss et al., 2016), many practices are structured by technological features and computational algorithms of the platforms. For example, automated dispatch systems use genetic or optimization algorithms and devices with built-in GPS to match drivers with passengers in real time based on geo-locations (Karande and Bogiri, 2015; Rawley and Simcoe, 2013). Fares and payment rates are set based on locations, times of the day (e.g., higher in rush hours), and the services requested (e.g., single ride or shared ride). In addition to real-time data, Uber assigns work to drivers and allows passengers to request services based on the historical data, namely the rating system on the platform (Ahmed et al., 2016).

Much previous work has investigated issues revolved around such computing systems and algorithms, and their influences on users. Automated dispatch systems may deploy drivers to move outside their familiar geographic areas (Hsiao et al., 2008). While this allows drivers to acquire information about some potential hotspots, it also demands drivers to develop temporal and spatial knowledge. Devices with GPS systems shape drivers' wayfinding and navigation skills and potentially change the social dynamics of the riding processes between drivers and passengers (Girardin and Blat, 2010; Hsiao et al., 2008). With their influences on practices and work revolved around the P2P platforms, the most prominent issue with these algorithms and systems is the lack of transparency to users (Lustig et al., 2016). Despite the invisibility and inaccessibility, users still have to make sense of how to interact with the systems in order to manage their work (Lee et al., 2015), rely on the digital infrastructure to quantify their work and develop their accountability using the rating system (Scott and Orlikowski, 2012), or deal with potential offline consequences like the uncertainty of finding next customer by taking request from the dispatch system (Ahmed et al. 2016). The lack of algorithmic transparency contributes to the large amount of emotional labor these workers must undertake to maintain their standing, this is a particularly problematic aspect of the nature of the rating systems for these platforms, particularly for Uber (Glöss et al., 2016; Raval and Dourish, 2016).

Algorithms and the computer systems that use them are designed to collect data to facilitate coordination or even prediction of human work, and are of course valued for their instrumental functions. Given these identified issues, computing systems and algorithms may not be posed as neutral and objective as they may seem (Kneese et al., 2014; Friedman, 1996). It is possible that the digital infrastructure imposes and renders biases, intentionally or unintentionally, against users (Wagenknecht et al., 2016).

In this study, we complement prior work by exploring and identifying how biases play out on the Uber platform. We examine the role of the platform and expand previous frameworks on biases in computer systems (Friedman, 1996). We draw from the accounts of biases provided by drivers in their discussions with other drivers, using Uber as our target platform, in an attempt to begin to flesh out and draw a picture around this issue, from at least one stakeholder's perspective.

## Biases in digital workplaces and computer systems

Drawing from Friedman (1996), we define bias as having a moral import that can be drawn from, and that for a system to exhibit bias it has to "systematically and unfairly discriminate against certain individuals or groups of individuals in favor of others" (Friedman, 1996, pg.332). Friedman (1996) outlined three types of bias that occur in computer systems, which are, *preexisting*, *technical* and *emergent*. More generally, biases usually refer to stereotypical generalizations based on sociodemographic or physical characteristics about certain groups that are assigned to the individual group members. Previous research reported gender biases (Heilman, 2012), ageism (Rupp et al., 2006), racial biases (Rosette et al., 2008), or weight bias (Rudolph et al., 2009) at traditional workplaces. These biases are associated with inequality in employment decisions, career advancement, performance expectation, workload, overall evaluations, etc.

While these biases are prevalent in physical workplace because the characteristics and attributions are visible and obvious to elicit implicit or explicit biases, they do not disappear even if the work is digitally mediated. Research has also reported that biases are similarly taking place on technological platforms. For example, workers on TaskRabbit used geolocations to evaluate whether to accept a task and were found to avoid distant and less well-to-do areas (Thebault-Spieker et al., 2015). On the other hand, clients may also choose workers from these P2P platforms based on their gender and race no matter if the tasks are completed in physical or virtual contexts (Hannák et al., 2017). Workers have to have adequate equipment like bank accounts, smartphone with built-in GPS or a fancy car in the case of Uber Black, to be able to provide services (Kasera et al., 2016).

Compounding these biases rendered by socio-demographical and physical factors, we argue that on the digitally mediated workplace, these biases could potentially be reinforced and propagated by the digital infrastructure.

The rating system on Uber represents a record of drivers' work performance and is used to evaluate their eligibility to receive service requests. However, there is no clear metric, such as driving skills, safety concerns, or decision-making strategies about picking up routes, as to how the performance is evaluated. Instead, drivers may have to engage in "emotional labor," in which they need to quickly build "micro-relationships" that make passengers feel good so as to get good ratings (Nardi, 2015; Rogers, 2015; Mcgregor et al.; Raval and Dourish, 2016; Rosenblat et al., 2016). Such emotional labor is easily influenced by random factors and the efficacy and accuracy of the rating system may benefit from a more holistic evaluation (Lee et al., 2015).

In addition, while racial and gender biases are suggested to be mitigated through Uber's matching algorithm, Mcgregor et al. (2017) pointed out that the algorithm actually denies users ability to choose their desirable drivers or passengers and therefore deepens the negative effect of expected homophily for both drivers and passengers. Instead, the consequence may be a lowered rating as opposed to avoidance. On the Uber platform, drivers usually have to respond to requests within 15 seconds without knowing the destination and expected fare in order to avoid deactivation from the platform. Uber drivers often do not have sufficient time for decision-making (Rosenblat and Stark, 2016) and have to deal with offline consequences reinforced by the platforms (Ahmed et al., 2016).

In our study, we explore several different occurrences of biased practices and judgements that are either enabled by the digital infrastructure or rooted in an aspect of it.

# Method

In investigating if and how Uber drivers discuss bias in the workplace, we borrowed heavily from the approach taken by Martin et al. (2014) in their study of Turkers' issues and concerns. We focused on the most popular forum for Uber drivers, UberPeople[1], a forum run by drivers for drivers. The primary way that we differ from Martin et al. (2014), is that, in this paper, we discuss a specific topic and do not report *all* of the topics that emerged from our study. We found that bias, while not always explicitly discussed, was a recurring theme and an important and influential topic; in fact, forum members clearly saw bias as related to the most popular topics in the forum. Among the most discussed topics such as transparency, algorithmic management, earnings and expenses, etc., bias happened along with, and as a result of these topics. Therefore, in order to understand the broader topics and concerns of drivers, it is critical to understand how bias plays into these different functionalities. We took an exploratory approach to our investigation around bias in the workplace, looking at all forms and instances, e.g. not just biases on the part of passengers, but also biases expressed by the drivers on the forums. Forum members were not aware that our study was being undertaken, we believe there are no ethical implications as these posts are made in a public place and no special privilege or access is needed to read this content. Our study was deemed exempt by our Institutional Review Board, as the forum was publicly available and open in nature.

The current users of UberPeople are from major cities around the world with most of active members located within the U.S. The forum is divided into 22 different sections, and the sections that we primarily draw from are: *Advice*, *Stories*, *People*, and *Complaints*. The *Advice* section is the most active section, closely followed by the *Complaint* section, the other sections *Stories* and *People*, have significantly less activity. While we read all of the sections systematically, the primary source of the content in this paper are from the *Complaints* section.

---

[1] https://uberpeople.net

For two months, we have been collecting content from the forum and gathering threads posted between January 2015 - February 2017. In selecting these threads, the authors of this paper read over various posts and discussed which threads involved discussions of bias. The threads that we draw from in this paper were selected because they represent a range of practices and scenarios in which biases occur in the workplace. For each of these threads, we analyzed every post in the thread (even though the majority of posts in a thread are quite terse) as a group and performed a thematic analysis. In some of these threads, the context of the thread was the topic of bias, but for the majority, the discussion of bias followed as an explanatory feature of the phenomenon being discussed (primarily either the rating system, the assignment of riders, or emergent practices).

To gauge how broadly felt the content of the different posts were, we looked at the responses of the community. For instance, if a user wrote a post making an uncommon, potentially outrageous claim, then the community would respond in kind. That said, expressing outrage at a claim of bias is not uncommon and was not necessarily an exclusion criteria. However, if the community is supportive and is in agreement this is a strong sign that a belief or experience is generally accepted by the community. For any threads that contains a mix of opinions on the part of the forum users, we situate the quote within the context of the discussion. All the selected posts and threads are categorized as being rooted in either a lack of transparency or lack of recourse. While presenting the different themes that emerged we make note of whether or not these are biases impacting drivers or passengers.

The categorization that we present in our findings is a result of our thematic analysis of the exemplars of bias on the Uber platform. For each quotation, we have anonymized the user, and each user is labeled with an F and a unique number.

# Findings

In our reading of the UberPeople forum, a number of themes emerged from our analysis of the discussion of biases on Uber: some biases seemed to be built into the platform itself, mapping to a *technical* bias (Friedman, 1996, pg.334); other preexisting personal biases were enabled or amplified by the platform; and some biases were in response to aspects of system use. That is, there are some biases that are seen as inherent in the design of the Uber marketplace and tool. Meaning that, there are other biases that are propagated or supported by the system unwittingly, as they clearly preexist and originate from one of the stakeholders and are clearly directed at another specific stakeholder. The platform as a vehicle for biases goes somewhat beyond the initial framework of Friedman (1996), which focused more on biases manifested in the *design* of the system and less in the usage. Somewhat surprisingly to us, we encountered a diverse set of biases in the forum, that is, while we expected to – and did – see biases that impacted the drivers (who after all were the primary users of the forum), we also saw discussions about biases aimed towards passengers by both the drivers and the platform structure. During our analysis, we saw two main roots to the perception or practice of biases: the lack of transparency in the system's policies and algorithms, which manifested mostly in the rating system; and the lack of recourse: there was no clear way to take recourse against what drivers saw as biased judgements, so they developed strategies, which contained biases.

## Biases Rooted in a Lack of Transparency

One of the frustrations that drivers had with Uber's rating system is that it is not transparent with respect to passengers' ratings, specifically regarding what the complaint was and who made it. Drivers especially concerned when they had received low ratings. In a thread where drivers discuss their low ratings.

> *The reason why we need to know who rated to be able to fix any issue ... This system will make riders more accountable before they ruin someones life. - F1*

At times, this lack of transparency led drivers down a path of suspicion. As reported in previous work (Raval and Dourish, 2016; Glöss et al., 2016), it is hard for the drivers to know what exactly they did to deserve a poor rating and they began to speculate about a variety of reasons. When drivers belong to a minority and are receiving low ratings for reasons that are unknown to them, they begin to speculate – with ample reasons at times – that it is related to a particular bias on account of the passenger.

### Biases at Play in Ratings

Drivers are clearly aware of the possibility for biased ratings, as well as the inability to know whether or not bias has influenced their ratings. Particularly, drivers that belong to a minority are concerned that the biases of their passengers may be impacting their rating. That said, all drivers speculated that this might be a problem. One new driver, who belonged to a minority, believed that they were suffering from biased ratings, which was particularly problematic as they just started and were in danger of being deactivated.

> *This is my 4th day driving. My rating now stands at 4.64... I just can't figure out why my rating are borderline deactivation level. This is crazy. I'm curious, especially to hear from other young(ish) black male drivers if they are constantly on the*

> *borderline as well. I hate even having to bring up this topic, but honestly I don't know what else I could even be doing to bring my rating up. - F2*

This particular driver was trying to figure out ways to raise the score before s/he got deactivated, and asked other drivers how they brought their ratings up. Responding to this driver's post, someone agreed with the speculation of bias.

> *If I were black and got deactivated I'd be screaming from the hilltops about racism. It's probably THE best argument against the rating system there is... Ageism is absolutely a factor too. But if you are an older black male I would say it's worse... But the bottom line is the ratings are unfairly applied. It probably depends on the area and the demographics of the customer base as to HOW they are unfairly applied. But anyone who thinks race isn't a factor (and ageism and sexism) in any system is deluded. - F3*

Conversations around biases, particularly racism, seem to become contentious fairly quickly on the forum (similar to other venues). When the issue is specifically called out by a user, passionate voices fall on both sides of the issue. Along the same conversation of the minority driver, some minimize and deride the claims of bias:

> *Every bad thing in your life that happens to you is racially motivated. "The man" is out to get you. - F4*

Others provide support and counter other members to defend the original poster:

> *You can talk all the sh!t you like, I am a black man in America, I see, hear and experience racism on a weekly basis. - F5*

Clearly, racial bias is an issue on which the community has very different opinions. However, racial bias was not the only type of bias that concerned drivers. There were other biases related to English fluency that one driver claimed to have noticed.

> *I've noticed a number of posts by poor-English speakers about bad ratings. That's probably one of the most difficult biases to overcome. - F6*

One user hypothesized that all manner of biases are probably at play in the rating system.

> *Of course the crowd-sourced rating system is racist. Probably sexist and ageist too. Ugly people get lower ratings than attractive people too. - F7*

It seems clear that the lack of transparency behind the reasoning for passengers' ratings is opening the door to biased ratings that are unfettered by the system. At the very least, this lack of accountability, mostly due to the anonymity of the ratings/complaints, in the ratings system is leading to a lot of suspicion.

Assignments of Passengers

The general lack of transparency in many of Uber's functionalities caused drivers to be suspicious that the algorithms by which passengers were assigned to them included hidden biases. Drivers speculated that Uber assigns certain types of passengers or passengers from certain types of areas to certain types of drivers:

> *I think as much as possible Uber tries to send us black drivers into the "hood".... To pick up black passengers.... This morning I was at the air port the 3rd one to go out....when I get a ping...I look at my phone, and see the pax is 25 min away and has a very ethnic specific name - F10*

Although this was met with skepticism from other drivers, one of the most prevalent strategies that other drivers provided as a solution was for the driver to be more selective about what types of neighborhoods or distances that they traveled for their passengers. Meaning that one of the most suggested strategies to deal with the biases, is to enact them proactively.

## Strategies in Response to Perceived Bias

While there is evidence on the forums that drivers at least perceive that they are impacted by the biases of passengers, there is also clear evidence of the various strategies that drivers had developed in response. In fact, the biases that we saw on the part of the drivers were surprisingly rooted in practices that drivers had enacted as a strategic response to the perception of passenger biases.

Ignore and Accept

One of the more innocuous strategies, at least with regards to how it impacted the passengers, was to just tolerate the bias as a part of doing business. They advised not to worry about it as cases of bias are absorbed by the majority of good, decent passengers and as time when on these incidents had less and less impact on their overall rating.

> *Seriously, do not worry about your rating this early in the game. I get the exact same BS feedback you got at 4.92 ratings after 500 plus rides. - F11*

However, to simply tolerate this intolerance is anathema to the zero-tolerance policy to which Uber subscribes[2], and certainly is not part of the type of equitable workplace that we should expect. That is, it is not an innocuous strategy in regards to the drivers.

Retaliation and Protest

In one case, a driver had become frustrated with receiving poor ratings for inscrutable reasons, so they decided to take a protest action. Whenever they received a poor rating, they gave each and every passenger they gave a ride to that day a poor rating.

> *Ok. So since Uber doesnt let us know who give us a bad rating and leave us guessing. I decided to punish all riders of the day if my rating goes down .01 point. ... I think we have the right to know who rate us bad and the reason. Otherwise i will use this method. I know it wont matter. But when the rider check their ratings they will see how it dipped down too. - F1*

In another thread discussing the effect of biased ratings on the drivers, the conversation turned towards speculation about 'certain areas' and 'stupid biases' being the source of poor ratings. In this case, the reaction to the discussion of biases was to conjure additional biases where the driver themselves implement biased practices. One user had taken a similarly oppositional practice of awarding high ratings only to exceptional passengers and to just accept that 'certain areas' are problematic.

> *Im done worrying about riders so much. If you work around certain areas. Youll realize your rating drops even if you keep the cleanest car and is the best driver. Now the pax needs to amuse me to get over 4 stars. Stupid Biases and complexes really get in the way. - F14*

Avoidance of Demographic Groups

The instances of driver bias towards passengers mostly happened in how the drivers tried to avoid certain areas or types of passengers. One example, is a driver who, after a bad experience with passengers from the Black Entertainment Television awards, experienced a dip in their rating and came to this conclusion:

---

[2] https://www.uber.com/legal/policies/zero-tolerance-policy/en/

> *I'm not ignorant of the racial tensions in this country right now. I'm sure there's some real animosity. I think there's something about Rap too that brings out the hate. Now when I see a group of black guys I'm automatically going to just hit cancel. I hate saying that too because I love my black friends but what are you going to do. - F9*

In this same thread, other drivers provided numerous counter examples where they had positive experiences with African American passengers. Clearly, there is the potential for drivers' biases to impact passengers' ability to procure a ride.

A different driver had another set of much more blatantly racist complaints about a different group of riders, framing them as others that even inhabit a different world of sorts.

> *1 They do not know this is a ride-sharing. They treat you like a low-educated, no-skill cab driver. 2 They intentionally make you wait for up to 5 minutes 3 They ask you drive up to the front door even they live in an apartment complex...4 Most of them have very strong body odors... 5 Most of their rides are a $4 trip including pick up from or go to the Indian grocery store or Indian restaurant...7 They never tip...8 They gave you wrong directions and blame you taking the longest route from point A to B. 9 They give you lower rating too. In their world, a 5star is impossible and never exists. - F8*

The avoidance strategies made available by Uber's cancellation functionality – which lets drivers cancel rides and suffer few, if any, consequences – were sometimes used by 'experienced' drivers to avoid passengers and areas. These strategies do have a negative impact on the passengers, which can be seen in one of the rare instances of a passenger posting to the forum.

> *This guy wasted my time (which apparently was very precious in that span), didn't answer my calls, THEN had the nerve to charge me a cancellation fee! Isn't there some way to rate this guy as unprofessional? I have his ID number. - F12*

This passenger was canceled by the driver on a day with severe weather. Due to the app system design, the passenger was charged a fee while his/her trip was canceled. This shows that there is at least a reciprocal avenue through which passengers can also be impacted by drivers' biases.

# Discussion

Friedman (1996) outlined a framework for analyzing bias in computer systems and stressed that freedom from bias should be among the criteria by which computer systems are judged as effective and appropriate by society. In their framework, the major categories were *preexisting*, *technical*, and *emergent*. In this work Friedman (1996) looked primarily at how biases impact the *design* of a system. One example of this outlook is *preexisting* bias, which is divided into two subcategories (Friedman, 1996, pg.334): *Individual*, which are biases that impact an individual system designer; and *Societal*, where larger, more cultural biases impact the system design. However, an additional aspect of bias that must be taken into account, is how a computer system can be an instrument of vehicle of bias. While this is related to Friedman's (1996) concept of the formalization of human constructs (Friedman, 1996, pg.334), we feel that it must be more explicitly dealt with when our systems are increasingly more *socio-technical*.

Rosenblat et al. (2016) used the Uber rating system as an example of a system that can be a "vehicle for bias." In our analysis, Uber drivers are also concerned about the possibility of their ratings being impacted by passenger biases. What surprised us, is that when we set out to more explicitly look at driver discussions around bias, we expected the drivers to be discussing the impacts of biases on themselves. What we did not expect, was the candor with which the drivers discussed their own biases (primarily as a response to perceived passenger bias) and how these biases impacted passengers. One forum member felt that the various avoidance strategies that drivers used were being reinforced by the various pricing strategies that Uber employs.

> *Uber has brought back redlining[3] with its boost incentives. It is subsidizing the rides of the well off, mostly white riders on the west side and leaving minorities and lower income residents in Central LA and South LA with fewer drivers. Uber, ..., are the ones responsible for ride share redlining ... - F13*

Not only do we see the importance of providing freedom from biased interactions for all stakeholders in a system, we also see that these biased interactions serve as the root of further biased interactions. Clearly, we must design socio-technical systems plainly considering how they might be used as a vehicle and proliferator of biases. Not to do so validates and expands our existing biases.

---

[3] Redlining is a practice that originates in more traditional taxi companies, where the companies refused fares from low-income communities. This practice of taxi companies was dealt with via legislation, but now seems to be reemerging on Uber.

## Transparency

To us, it seemed clear that the main root of biases in the Uber app was a lack of transparency in how the system functioned. This is true in two regards: First, a general lack of accountability is a direct effect of a lack of transparency, which frees individuals to express their existing biases. Second, the lack of transparency also breeds suspicion, which breeds additional biases or at least is a method for reinforcing existing biases.

Rosenblat et al. (2016) discussed how rating systems can serve as vehicle of bias, we contend that a key contributor to biased ratings is a lack of accountability caused by the lack of transparency. When biases are more apparent and obvious, the public is able to apply pressure to companies and cause them to take action. Such was the case when a Raleigh, NC same-sex couple was kicked out of an Uber driver's car, their story was covered in the media and discussed later in the forum with mixed voices. Subsequently, Uber released a statement condemning the interaction and blocked the driver from giving future rides on Uber. However, the small instances of bias that we have seen evidence of, be it by either drivers or passengers, are much more difficult to trace and take action on.

These circumstances lead some drivers down – sometimes perhaps further down – the path of bias. At times, some drivers' reaction is to exercise their own biases, sometimes perhaps they are not quite aware of what they are doing. This is perhaps a predictable reaction to a system that is both high-stakes, in that drivers' access to the market will be shutdown if their rating drops below the acceptable rate, and obscure, in that drivers have little knowledge about how rides are distributed and why or even when they were given poor ratings.

## Design Implications

There are two preliminary design implications that come from our findings. First, we argue for a higher degree of transparency behind user ratings of each other. Perhaps, protecting the various stakeholders from awkward situations by depersonalizing interactions through digital mediation is not the right way to go. There almost seems to be an inclination to bring the anonymity of the online world to our face-to-face interactions. Maybe, uncomfortable situations can serve a regulating purpose in socio-technical system. Perhaps, if drivers/passengers would like to give one another a poor rating or deny them a ride, this should be visible on the platform. Giving individuals protection from the consequences of their actions may not lead to more responsible behaviors.

On the more proactive side, there is a possibility that rating systems (like the one Uber users) can better leverage the various data that they are gathering about ratings and interactions. For instance, Uber can keep track of each passenger's reactions to different demographics and use this information to reduce the weight of that person's ratings if s/he shows systemic bias. Additionally, the passengers could be confronted with this perceived bias, as it may be implicit and not realized, so that they can act to remedy their own bias or at least know that 'someone' has noticed. If the biased interactions continue, more formal action can be taken by the platform, such as denying access.

# Limitations

Our preliminary study has obviously limitations in the length of time that we have been collecting data and the breadth of data that we have included. That said, we feel that we have several concrete examples of a phenomenon that is rarely discussed, which map to the bias that other researchers have reported on these platforms. We have also begun to identify some of the strategies that drivers have taken in response to perceived bias.